\documentclass[twocolumn,showpacs,preprintnumbers,amsmath,amssymb]{revtex4}
\usepackage{graphicx}
\usepackage{dcolumn}
\usepackage{bm}

\begin{document}


\title{CeFePO: A Heavy Fermion Metal with Ferromagnetic Correlations}

\author{E.M.~Br\"uning}
\author{C.~Krellner}
\author{M.~Baenitz}
\author{A.~Jesche}
\author{F.~Steglich}
\author{C.~Geibel}

\affiliation{Max Planck Institute for Chemical Physics of Solids,
D-01187 Dresden, Germany}

\date{\today}
\begin{abstract}

The ground state properties of CeFePO, a homologue of the new high
temperature superconductors $R$Fe$Pn$O$_{1-x}$F$_x$, were studied by
means of susceptibility, specific heat, resistivity, and NMR
measurements on polycrystals. All the results demonstrate that this
compound is a magnetically non-ordered heavy fermion metal with a
Kondo temperature $T_K \sim 10$~K, a Sommerfeld coefficient $\gamma
= 700$~mJ/molK$^2$ and a mass enhancement factor of the order of 50.
Analysis of the susceptibility data and of the spin relaxation time
indicate that the strong electronic correlation effects originate
from the Ce-4$f$ electrons rather than from Fe-3$d$ electrons. An
enhanced Sommerfeld-Wilson ratio $R = 5.5$ as well as a Korringa
product $S_0$/$T_1TK^2 \simeq 0.065$ well below 1 indicate the
presence of ferromagnetic correlations. Therefore, CeFePO appears to
be on the non-magnetic side of a ferromagnetic instability.

\end{abstract}



\maketitle

The compounds $RTPn$O ($R$: rare earth, $T$: transition metal, $Pn$:
P or As) have started to attract considerable attention because of
the recent discovery of superconductivity with a transition
temperature $T_c$ exceeding 50~K in the $R$FeAsO$_{1-x}$F$_x$ series
\cite{Kamihara2008,XHChen2008,Ren2008,ChenGF2008}, being the highest
$T_c$ except for cuprate systems. This high $T_c$ superconducting
state appears to be of unconventional nature and related to the
disappearance of a Fe-based spin density wave. While recent reports
focus on the importance of electronic correlation effects due to
3$d$-electrons close to a magnetic state, the homologous compounds
with $R$ = Ce are attractive candidates for 4$f$-electron induced
strong correlation effects. Thus last year, we presented a detailed
study of the properties of CeRuPO and CeOsPO
\cite{Krellner2007,Krellner2008,Krellner2008b} and demonstrated that
the former one is a rare example for a ferromagnetic (FM) Kondo
lattice with a FM Ce ordering temperature $T_C = $15~K and a Kondo
temperature $T_K \sim 10$~K, while the latter one shows
antiferromagnetic (AFM) order of stable trivalent Ce below $T_N$ =
4.5~K. More recently, superconductivity at 41~K was also reported in
CeFeAsO$_{1-x}$F$_x$ \cite{ChenGF2008}, but the relevance of
4$f$-correlations could not yet be analyzed because of the very
limited amount of experimental data. In this report we present a
complete study of the basic physical properties of CeFePO using
susceptibility $\chi(T)$, specific heat $C(T)$, and resistivity
$\rho (T)$ measurements as well as NMR as a local probe. The
analysis of our results indicates that CeFePO is a magnetically
non-ordered heavy fermion metal with strong 4$f$-based electronic
correlations and FM fluctuations, while Fe correlations seem to be
of minor importance. After submission of this paper, a very precise
investigation of LaFePO was reported by T.M.~McQueen {\it et. al.}
\cite{McQueen2008}. These authors conclude that LaFePO is a
non-magnetic metal with only weak exchange enhancement from spin
fluctuations.

Polycrystalline samples were synthesized using a Sn-flux method in
evacuated quartz tubes as described in
\cite{Krellner2007,Kanatzidis2005}. Several powder X-ray diffraction 
patterns recorded on a Stoe diffractometer in transmission mode
using monochromated Cu-K$_{\alpha}$ radiation ($\lambda =
1.5406$~\AA) confirmed the \textit{P4/nmm} structure type and the
formation of single phase CeFePO. The lattice parameters
$a=3.919(3)$~\AA $ $ and $c=8.330(5)$~\AA $ $ refined by simple
least squares fitting were found to be in good agreement with the
reported structure data \cite{Zimmer1995}. Susceptibility
measurements were performed in a commercial Quantum Design (QD)
magnetic property measurement system MPMS. $\rho(T)$ was determined
down to 0.4~K using a standard AC four-probe geometry in a QD
physical property measurement system PPMS. The PPMS was also used to
measure the specific heat $C(T)$ with a standard heat-pulse
relaxation technique. NMR measurements are carried out by monitoring
the spin-echo intensity as a function of applied field at a fixed
frequency (12.1~MHz, 27~MHz, 76.4~MHz, and 130~MHz). Shift values
are calculated from the resonance field $H^*$ by $K =
(H_{L}-H^{*})/H^{*}$ whereas the Larmor field $H_{L}$ is given by
using H$_3$PO$_4$ with $^{31}K = 0$ as a reference compound.
Relaxation rates $^{31}(1/T_1)(T)$ are determined by using the
saturation recovery method.

\begin{figure}
\includegraphics[width=8cm]{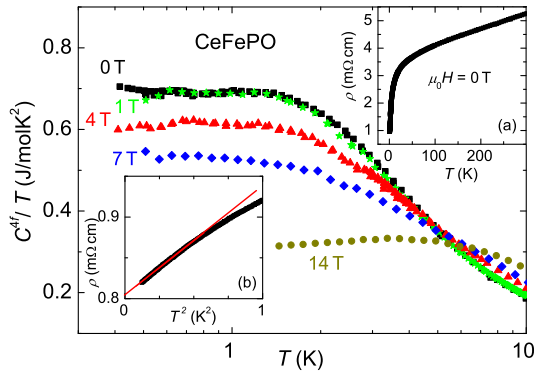}
 \caption{(Color online) 4$f$-increment to the specific heat of CeFePO
plotted as $C^{4f}/T$ vs $T$  in different magnetic fields. Inset (a)
shows $\rho(T)$ and (b) $\rho(T^2)$.}
\label{FigCTvT}
\end{figure}

\begin{figure}
\includegraphics[width=8cm]{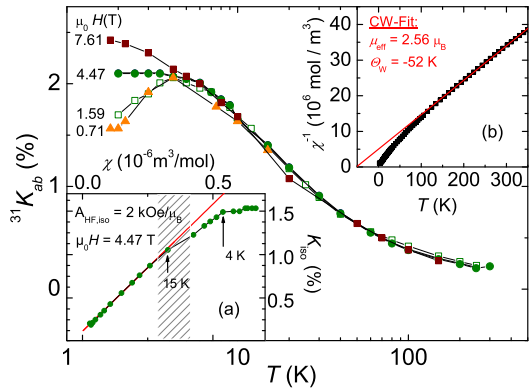}
 \caption{(Color online) Shift component $^{31}K_{ab}$ as a function of
temperature in different magnetic fields as indicated (Field values were
calculated from NMR frequency using ($^{31}\gamma_{\rm nmr}
/2\pi) $=17.10~MHz/T). Inset (a) shows the isotropic shift
$^{31}K_{\rm{iso}}=(2K_{ab}/3)+(K_{c}/3)$ versus isotropic (bulk)
susceptibility from MPMS measurements (5~T). Inset (b) shows $\chi^{-1}(T)$,
solid line corresponds to a Curie-Weiss (CW) fit.}
\label{Kab+Chi}
\end{figure}

We start by briefly presenting the results of $\chi(T)$ of CeFePO.
Below 350~K, $\chi(T)$ of CeFePO increases with decreasing
temperature following a Curie-Weiss law down to 100~K
(Fig.~\ref{Kab+Chi} inset (b)). The effective moment $\mu_{\rm eff}$
= 2.56~$\mu_B$, being almost identical to the expected value
$\mu_{\rm eff}$ = 2.54~$\mu_B$ for a free trivalent Ce ion, and the
moderate Weiss temperature $\theta_W$ = -52~K indicate a nearly
stable trivalent Ce state and the absence of a significant
contribution of a paramagnetic Fe-moment in this temperature range.
Below 100~K, $\chi(T)$ deviates from the Curie-Weiss law, increasing
stronger with decreasing temperature than expected, which is
attributed to crystal electric field (CEF) effects. Below 10~K, the
measured bulk susceptibility becomes strongly field dependent,
showing a pronounced increase for low fields $\mu_0H < 1$~T, while
for larger fields $\chi(T)$ levels off at a constant value $\chi_0 =
660 \times 10^{-9}$m$^3$/mol. While such a field dependence is
usually attributed to paramagnetic defects or impurities, the NMR
results suggest this to be an intrinsic property of CeFePO (see
below). No anomaly corresponding to magnetic order could be resolved
at any field. Thus, a preliminary analysis of these susceptibility
data suggests a paramagnetic Kondo lattice. This conclusion is
confirmed by the results of the specific heat measurements, plotted
as $C^{4f}/T$ vs $T$ in Fig.~\ref{FigCTvT}. The $4f$-contribution
$C^{4f}$ to the specific heat was obtained by subtracting the
reported specific heat data of LaFePO \cite{McQueen2008}. Below
10~K, $C^{4f}/T$ increases logarithmically down to 1~K, where it
levels off at a constant value $\gamma = 700$~mJ/molK$^2$. No
anomaly indicating a phase transition can be resolved above 0.4~K.
Application of a small magnetic field $\mu_0H < 1$~T does not change
$C^{4f}(T)$, while a larger field leads to a smooth decrease of
$C^{4f}/T$ at low temperatures. The former result confirms the
absence of long-range magnetic order, while the latter one is
incompatible with spin-glass-type behavior. Therefore, the large
$C^{4f}/T$ value at low $T$ has to be attributed to heavy fermions
and $\gamma$ corresponds to a strongly enhanced Sommerfeld
coefficient in the electronic (heavy-fermion-based) specific heat.
Comparing this with the Sommerfeld coefficient of the nonmagnetic
LaFePO ($\gamma=12.5$~mJ/molK$^2$) \cite{McQueen2008}, one estimates
a low-temperature enhancement factor of the order of 50 for the
heavy quasiparticles in CeFePO. The magnetic entropy $S(T)$ was
calculated by integrating the measured $C^{4f}(T)/T$ over $T$ for
$T<10$~K. The entropy $S(T)$ reaches 0.5$R\ln$2 at 5~K, indicating a
Kondo temperature $T_K \sim 10$~K. The absolute values of $C(T)$,
its temperature dependence as well as the deduced Kondo temperature
are very close to those of the archetypical heavy fermion metal
CeCu$_2$Si$_2$ \cite{Steglich1979}. The temperature dependence of
the electrical resistivity $\rho(T)$ shown in the inset (a) of
Fig.~\ref{FigCTvT} presents the typical behavior of a Kondo lattice
system and thus confirms the above conclusion. Below 300~K,
$\rho(T)$ decreases almost linearly, before dropping rapidly below
30~K due to the onset of coherent Kondo scattering.
Below $T=1$~K (see inset (b)), $\rho(T)$ follows a $\rho\propto T^2$
dependence indicating the formation of a Landau-Fermi liquid ground
state in agreement with the specific-heat data. However, the
absolute value of $\rho(T)$ is rather high; very likely, this is not
intrinsic of CeFePO but due to the granularity of the
polycrystalline sample.

\begin{figure}
\includegraphics[width=8cm]{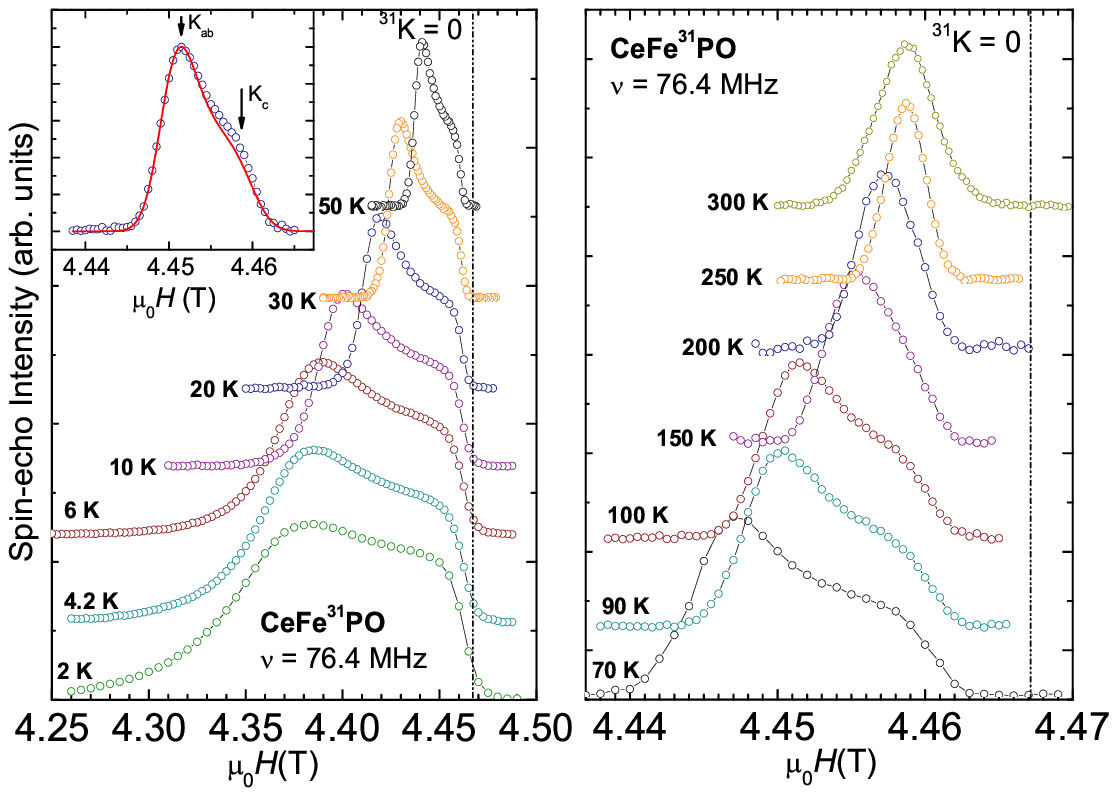}
\caption{(Color online) $^{31}$P field-sweep NMR spectra at $\nu = 76.4$ MHz
and different temperatures. Vertical dashed lines indicate the Larmor field
obtained from a reference compound. Inset shows a typical
powder spectrum, together with the simulation at $T = 50$~K (arrows
indicate resonance fields $H^{*}$ for the $H \perp c$ and $H
\parallel c$ direction).}
\label{Spectra}
\end{figure}

In order to get a deeper insight into the many-body phenomena in
this compound, we used $^{31}$P-NMR as a microscopic local probe.
A typical set of spectra at different temperatures is shown in
Fig.~\ref{Spectra} for a NMR frequency of 76.4~MHz (corresponding to
4.47~T). One single narrow $^{31}$P-NMR line as expected from the
crystal structure was found at room temperature (right panel of
Fig.~\ref{Spectra}). It develops strong anisotropy towards lower
temperatures (left panel of Fig.~\ref{Spectra}). The shape is
characteristic of a powder pattern from a spin $I=1/2$ nucleus in a
tetragonal symmetry. The broadened $^{31}$P-NMR lines could be
simulated consistently at all temperatures with shift-tensor
components $K_{ab}(T)$ and $K_{c}(T)$ corresponding to the $H \perp
c$ and $H \parallel c$ directions, respectively (inset in
Fig.~\ref{Spectra}). The relative shift values $^{31}K_{ab}(T)$ and
$^{31}K_{c}(T)$ are both positive, in contrast to the results
obtained for CeRuPO and CeOsPO \cite{Krellner2007}. While
$^{31}K_{c}(T)$ is almost independent of $T$, $^{31}K_{ab}(T)$
increases strongly with decreasing temperatures
(Fig.~\ref{Kab+Chi}).
Above 15~K, $^{31}K_{\rm{iso}}=(2K_{ab}/3)+(K_{c}/3)$ tracks the
bulk susceptibility $\chi(T)$ obtained by MPMS measurements at 5~T
(inset~(a)). Below 10~K, the line shape of the spectra becomes field
dependent. While at high fields ($\mu_0H > 6$~T) the line shape
observed at higher temperatures is maintained down to 2~K, reducing
the field below 4~T leads to a strong broadening of the contribution
related to $H \perp c$ and to a shift of its maximum to lower $K$
values. In contrast, the contribution connected with $H \parallel c$
is not affected. Since foreign phases or impurities usually do not
contribute to the NMR signal \cite{Dormann1991}, and since
impurities would not show such an anisotropic behavior, this
broadening cannot be explained by disorder or impurities. Such a
broadening is typical for the onset of short-range ferromagnetic
correlations. These correlations are likely to be responsible for
the field dependent low $T$ increase in $\chi(T)$, too. Since only
the component $H \perp c$ is affected in the NMR spectra, only the
basal-plane components of the magnetic moments start to be
ferromagnetic correlated. This strong anisotropy supports that these
correlations originate from the Ce-moments. A fit of the whole
spectra still lead to a reasonably well defined $K_{ab}$ value, but
the resulting $K$ is now field dependent. At high fields, $K_{ab}$
increases down to the lowest $T$, while at low fields it shows a
clear maximum. This specific field dependence appears to indicate
that CeFePO is close to a
ferromagnetic instability.

\begin{figure}
\includegraphics[width=8cm]{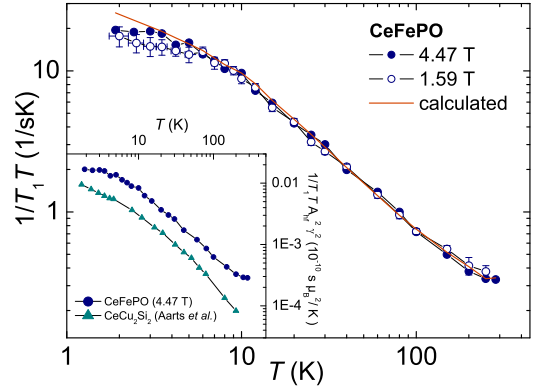}
 \caption{(Color online) $^{31}(1/T_1T)$ as a function of temperature for
different fields as indicated. Solid line represents the calculation
described in the main text. Inset shows the comparison with
$^{29}$Si-NMR ($I=1/2$) results on CeCu$_{2}$Si$_{2}$ (data taken from Ref. \cite{Aarts1983}). $(1/T_1T)$
is normalized by $(A_{\rm hf, iso} \gamma_{\rm nmr})^{2}$.}
\label{T1}
\end{figure}

Finally, we present spin-lattice-relaxation ($T_1$) data on CeFePO.
$T_{1}$ measurements were performed as a function of temperature at
two different frequencies, 27~MHz and 76.4~MHz (corresponding to
1.59~T and 4.47~T, respectively), at the $H \perp c$ position of the
anisotropic NMR line (left arrow in inset of Fig.~\ref{Kab+Chi}. The
nuclear magnetization recovery curves $M(t)$ could be fitted at any
temperature and field with a standard single exponential function.
$^{31}(1/T_1T)$ plotted in Fig.~\ref{T1} increases monotonously over
two orders of magnitude upon cooling down from 300~K to 4~K. At
lower temperatures, $^{31}(T_1T)^{-1}$ stays constant at about
20~(sK)$^{-1}$. Since NMR data on the reference compound LaFePO are
presently not available, we note that the homologue LaRuPO has a
$^{31}(T_1T)^{-1}$ value $\simeq 1$~(sK)$^{-1}$, more than one order
of magnitude smaller than that of CeFePO. Even $1/T_1T$ of $^{75}$As
in doped LaFeAO$_{1-x}$F$_x$ \cite{Nakai2008,Grafe2008} is smaller
than that of $^{31}$P in CeFePO indicating that the relaxation is
dominated by the 4$f$ contribution. No significant frequency or
field dependence of $T_1$ could be detected in CeFePO, in contrast
to the behavior of $K(T)$. In Kondo lattice systems, $1/T_1T$ is
dominated by the contribution of the 4$f$ electrons, which is within
simple approximations proportional to the ratio of the static
susceptibility $\chi(T)$ and the dynamical relaxation rate
$\Gamma(T)$ of the 4$f$ electrons, $1/T_1T \propto
\chi(T)/\Gamma(T)$
\cite{Nakamura1996,Buttgen1996,Kuramoto2000,MacLaughlin1989}.
Further on, for a Kondo system one may expect $\Gamma \propto
\sqrt{T}$ for $T > T_K$ \cite{Cox1985}, but merging into a constant
value $\propto T_K$ at low $T$. Since for $T > 10$~K, $\chi(T)$ and
$K(T)$ are proportional, we calculated the temperature dependence of
$1/T_1T$ using $1/T_1T = a \cdot K(T)/\sqrt{T}$, where $a$ is just a
$T$-independent parameter to scale the whole curve to the absolute
experimental values. We added a small $T$-independent contribution
$(T_1T)^{-1} = 0.12$~(sK)$^{-1}$ to account for the contribution of
the conduction electrons at high $T$, where the contribution of the
4$f$ electrons is vanishing. The results of this calculation shown
as the solid line in Fig.~\ref{T1}, agrees very well with the data
from 300~K down to 10~K, showing that the dynamical relaxation rate
$\Gamma$ indeed follows the expected $\sqrt{T}$ behavior. The
deviation below 10~K is as expected, since $\Gamma(T)$ should merge
in a constant value instead of decreasing further. Further on, we
compare in the inset of Fig.~\ref{T1} $^{31}(1/T_1T)$ of CeFePO with
$^{29}$Si-NMR ($I=1/2$) results of the prototypical heavy fermion
metal CeCu$_{2}$Si$_{2}$ \cite{Aarts1983} which has a similar Kondo
scale. For CeCu$_2$Si$_2$ we specifically chose the NMR results on
$^{29}$Si in order to compare data of two nuclei with $I$ = 1/2 and
used the results of Aarts \emph{et al.}, because these data were
obtained at a field comparable to our measurements at 76.4~MHz
(4.47~T). To account for different NMR nuclei and hyperfine
couplings, $(1/T_1T)$ values are divided by $(A_{\rm hf, iso}\cdot
\gamma_{\rm nmr})^{2}$. The comparison can then be performed on an
absolute scale without
having introduced model dependent parameters. Both curves share the 
same overall temperature dependence but CeFePO shows larger absolute
values. The very similar $T$-dependences $\propto \chi(T)/ \sqrt{T}$
found for CeFePO and CeCu$_2$Si$_2$ hint at similar spin-fluctuation
relaxation mechanisms in both compounds and confirm that the strong
correlation effects in CeFePO originate from the Ce-4$f$ electrons
rather than from Fe-3$d$ electrons. Further on, since this
similarity is observed on the same absolute temperature scale, it
confirms comparable Kondo temperatures in both compounds. The larger
values of the normalized $1/T_1T$ in CeFePO perfectly agree with the
higher susceptibility. For CeFePO, $\chi$(2 K) saturates at $660
\times 10^{-9}$~m$^{3}$/mol in high fields whereas for
CeCu$_{2}$Si$_{2}$ $\chi$(2 K) $\approx 100\times
10^{-9}$~m$^{3}$/mol was published \cite{Sales1976}. Since both
compounds have almost the same Sommerfeld coefficient in their
electronic specific-heat, a larger susceptibility implies an
enhanced Sommerfeld-Wilson ratio $R = \chi/\gamma \cdot R_0, R_0 =
\pi^2k_B^2/ (\mu_0 \mu_{\rm eff}^2$) in CeFePO. With $\mu_{\rm eff}
= 1.73$~$\mu_B$ corresponding to an average value $\mu_{\rm
sat}=1~\mu_B$ expected for a Ce$^{3+}$ doublet ground state
\cite{Fischer1990}, we obtain $R = 5.5$, which highlights the
presence of dominating FM correlations in CeFePO. This scenario is
supported by the analysis of the Korringa product, given by $T_1TK^2
\equiv S_0 = \mu_{\rm eff}^2 / 3 \pi \hbar  k_B \gamma_n^2$.
Magnetic correlations modify this relation by introducing an
enhancement factor $K(\alpha)$, given by $T_1TK^2 = S_0/K(\alpha)$
where $K(\alpha) > 1$ and $0 < K(\alpha) < 1$ reflects the existence
of AFM and FM correlations among the quasiparticles, respectively.
For CeFePO, a value of $K(\alpha) \simeq 0.065$ is found at 2~K
(4.47~T), in fact indicating strongly dominating FM correlations. A
value $K(\alpha) \ll 1$ is also found in other heavy fermion metals
with dominating FM correlations like YbRh$_2$Si$_2$
\cite{Ishida2002} and CeRu$_2$Si$_2$ \cite{Kitaoka1985c}. In
contrast, for CeCu$_2$Si$_2$ with dominating AFM correlations,
$K(\alpha)> 1$ was determined \cite{Tou2005}.

To conclude, we prepared polycrystalline samples of CeFePO by a Sn
flux technique and determined its ground-state properties by means
of susceptibility, specific-heat, electrical resistivity and NMR
measurements. $\chi(T)$ follows a Curie-Weiss law at high
temperatures with an effective moment expected for free trivalent
Ce$^{3+}$ but levels off below 4~K at a constant, enhanced Pauli
susceptibility $\chi_0 = 660 \times 10^{-9}$~m$^3$/mol.
$C^{4f}(T)/T$ increases logarithmically below 10~K and also
saturates below 2~K at $\gamma = 700$~mJ/K$^2$mol. The temperature
dependence of $\rho(T)$ presents the typical behavior of a heavy
fermion metal with a strong decrease when cooling down to below 30~K
due to the onset of coherent Kondo scattering. In the NMR
measurements, $1/T_1T$ increases roughly with $\chi(T)/\sqrt{T}$
upon cooling from 300~K to 10~K, but saturates below 10~K at a
strongly enhanced Korringa term. Thus, CeFePO presents the typical
properties of a classical Ce-based heavy fermion metal. At low $T$,
we observe the behavior expected for a Landau-Fermi liquid: a
$T$-independent susceptibility $\chi_0$, a $T$-independent $C(T)/T$,
a $T$-independent $1/T_1T$ as well as a resistivity increment with
$T^2$. None of the investigated properties gives evidence for
magnetic ordering above 0.4~K. Therefore, CeFePO may be classified
as a paramagnetic heavy fermion metal. The absence of significant
Fe-contribution to the effective moment at elevated temperatures
indicates that the magnetism in CeFePO is dominated by the Ce-4$f$
electrons. Thus, the observed strong electronic correlation effects
originate from the Ce-4$f$ electrons rather than the Fe-3$d$
electrons. An enhanced Sommerfeld-Wilson ratio as well as a reduced
Korringa product highlight the presence of dominating FM
correlations. Therefore, CeFePO is likely to be located on the
non-magnetic side close to a FM instability.





\end{document}